\documentclass[twocolumn,runningheads,natbib]{svjour2}
\journalname{Astrophysics and Space Science}
\smartqed  
\usepackage{graphicx}
\usepackage{mathptmx}      

\begin{document}

\title{Isolated neutron stars: Magnetic fields, distances, and spectra}

\author{Marten H.\ van Kerkwijk \and David L.\ Kaplan}

\institute{M. H. van Kerkwijk\at
           Department of Astronomy \& Astrophysics\\
           University of Toronto\\
           60 Saint George Street\\
           Toronto, Ontario~~M5S 3H8\\
           Canada\\
           Tel.: +1-416-9467288\\
           Fax:  +1-416-9467287\\
           \email{mhvk@astro.utoronto.ca}
      \and D. L. Kaplan\at
           Kavli Institute for Astrophysics and Space Research\\
           Massachusetts Institute of Technology\\
           77 Massachusetts Ave, Room 37-664H\\
           Cambridge, MA 02139\\
           USA\\
	   Tel.: +1-617-2537294\\
	   Fax: +1-617-2530861\\
           \email{dlk@space.mit.edu}
 }

\date{Received: date / Accepted: date}

\maketitle

\begin{abstract}
We present timing measurements, astrometry, and high-resolution
spectra of a number of nearby, thermally emitting, isolated neutron
stars.  We use these to infer magnetic field strengths and distances,
but also encounter a number of puzzles.  We discuss three specific
ones in detail:
(i) For RX J0720.4$-$3125 and RX J1308.6+2127, the characteristic ages
are in excess of 1~Myr, while their temperatures and kinematic ages
indicate that they are much younger;
(ii) For RX J1856.5$-$3754, the brightness temperature for the optical
emission is in excess of that measured at X-ray wavelengths for
reasonable neutron-star radii; 
(iii) For RX J0720.4$-$3125, the spectrum changed from an initially 
featureless state to one with an absorption feature, yet there was
only a relatively small change in $T_{\rm eff}$.
Furthermore, we attempt to see whether the spectra of all seven
sourced, in six of which absorption features have now been found, can
be understood in the context of strongly magnetised hydrogen
atmospheres.  We find that the energies of the absorption features can
be reproduced, but that the featureless spectra of some sources,
especially the Wien-like high-energy tails, remain puzzling.
\end{abstract}

\section{Introduction}
\label{intro}
One of the great benefits of the {\em ROSAT} All-Sky Survey
\citep{rbs} is that is has provides an unbiased sample of all classes
of nearby neutron stars (limited only by their age and distribution of
the local interstellar medium).  Particularly interesting is the
discovery of the group of seven nearby, thermally emitting, isolated
neutron stars (INS; for a review, see Haberl, these proceedings).

The INS form the majority among the nearby neutron stars (typical
distances are less than $\sim\!500$~pc; \citealt{kvka02}; see also
Posselt, Popov, these proceedings), yet are atypical of the
neutron-star population represented by radio surveys: while pulsars
detected by their thermal emission all have normal periods of less
than a second, five out of the seven INS have periods about ten times
longer (the remaining two appear to have no pulsations despite
intensive searches; \citealt{rgs02,vkkd+04}).  A number of models ---
accretors \citep{w97}, middle-aged magnetars \citep{hk98,hh99},
long-period pulsars \citep{kkvkm02,zhc+02} --- have been suggested to
explain these objects.

A prime reason for studying the INS is the hope of constraining
fundamental physics at very high densities: neutron stars are natural
laboratories for quantum chromodynamics \citep{rho00}.  The overall
goal is to determine the masses and radii of a number of neutron stars
and hence constrain the equation of state (EOS) of ultra-dense matter
(\citealt{lp00}; Lattimer, these proceedings).  For the majority of
known neutron stars (i.e., radio pulsars), this is complicated by the
non-thermal emission that dominates the spectrum, but for the INS this
is not the case: the X-ray spectra show thermal emission only.  Hence,
much effort has been spent trying to derive constraints from the INS
\citep{bzn+01,bhn+03,dmd+02,pwl+02}.  The constraints have not been
very meaningful, however, because the data could not be interpreted
properly: they just do not fit any current realistic models
\citep{mzh03,ztd04}.

To make progress in understanding the thermal emission, we need first
to know the basic ingredients: the elemental abundances, the
temperature distribution, and the magnetic field strength.
Furthermore, to use the thermal emission to infer radii, we need
information about the distance.  Fortunately, observational clues are
now becoming available: broad absorption features at energies of
0.3--0.7~keV have been discovered in the spectra of six of the seven
INS (\citealt{hsh+03,hztb04,hmz+04,vkkd+04,zct+05}; see Haberl, these
proceedings), and, as described below, magnetic field strengths have
been inferred from timing solutions and new or improved parallaxes
have been measured.

The outline of this contribution is as follows.  First, in
\S\ref{sec:timing}, we present timing solutions for RX J0720.4$-$3125
and RX J1308.6+2127, and discuss the resulting estimates of the
magnetic field strengths and characteristic ages.  Next, in
\S\ref{sec:d}, we describe new parallax distance measurements for RX
J1856.5$-$3754 and RX J0720.4$-$3125.  For the former, these resolve
previous conflicting results, but also raise a puzzle: a rather large
radius or high brightness temperature inferred for the optical
emission.  In \S\ref{sec:spectra}, we turn to high-resolution X-ray
spectra, comparing spectra of RX J0720.4$-$3125, before and after its
spectral change, with those of RX J1308.6+2127.  In
\S\ref{sec:atmosphere}, we attempt to interpret the observations
assuming the sources have gaseous atmospheres, focussing on hydrogen,
but also briefly discussing the possibility of helium.  We summarise
and discuss future work in \S\ref{sec:discussion}.

From here on, we will refer to the INS in the text using abbreviated
names: J0420 for RX J0420.0$-$5022, J0720 for RX J0720.4$-$3125, J0806
for RX J0806.4$-$4123, J1308  for RX J1308.6+2127 = RBS 1223, J1605
for RX J1605.3+3249, J1856 for RX J1856.5$-$3754, J2143 for RX
J2143.0+0654 = RBS 1774 = RXS J214303.7+065419.

\begin{figure}
\begin{center}
\includegraphics[width=\hsize]{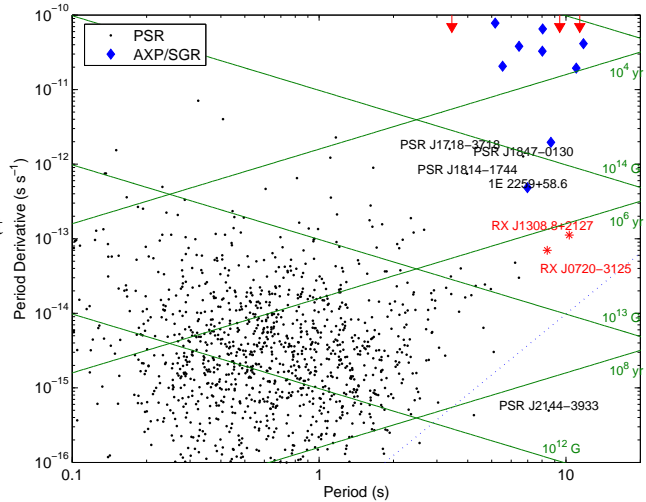}
\caption{$P$-$\dot P$ diagram, showing radio pulsars (points) and
  magnetars (diamonds); selected objects are labeled.  Also shown are
  the five INS with periodicities: RX J0720.4$-$3125 and RX
  J1308.6+2127 are shown by the stars, while RX J0420.0$-$5022, RX
  J0806.4$-$4123, and RX J2143.0+0654 are the arrows at the top (since
  $\dot P$ is unknown).  The diagonal lines show loci of constant
  dipole magnetic field and spin-down age, as labeled.}
\label{fig:ppdot}
\end{center}
\end{figure}

\section{Timing solutions}
\label{sec:timing}

Until recently, the typical magnetic field strength of the INS was
just a guess, with a wide range of possibilities
($10^{10}$--$10^{15}$~G) -- similar to the wide uncertainty in
magnetic field strength implied for the different demographic models
of the INS, although based on the long spin periods fields of a few
$10^{13}$~G were considered most likely.

We have improved upon this situation using dedicated timing
observations with {\em Chandra}, which, combined with {\em Chandra},
{\em XMM-Newton}, and {\em ROSAT} archival observations, allowed us to
determine phase-coherent timing solutions for J0720 and J1308
stretching back at least 5 years \citep{kvk05,kvk05b}.  Using the
periods and the period derivatives to place these objects on the
classic $P$-$\dot P$ diagram (Fig.~\ref{fig:ppdot}), one sees that
they are intermediate between radio pulsars and magnetars (in line
with the idea that they are long-period pulsars whose radio beams do
not cross our line of sight; \citealt{kkvkm02,zhc+02}).

From the solutions, assuming the sources spin down by magnetic dipole
radiation, one infers that they have similar magnetic fields:
$B=2.4\times10^{13}$~G and $3.4\times10^{13}$~G for J0720 and J1308,
respectively.  As will become clear below (\S\ref{sec:atmosphere}),
this agrees quite well with the magnetic field strengths inferred from
the absorption lines.

One also infers characteristic ages: $\tau_{\rm c}\equiv
P/2\dot{P}=1.9~$Myr for J0720 and 1.5~Myr for J1308.  These are
puzzling, since they are substantially longer than expected based on
cooling: From standard cooling curves \citep{plps04}, the observed
temperatures around 90~eV ($10^6$~K) correspond to ages of a few
$10^5~$yr.  Even if one takes into account that the black-body
temperature likely overestimates the effective temperature -- as is
clear from the fact that the extrapolation of a black-body fit to the
X-ray data underpredicts the optical -- one is very hard-pressed to
find an age in excess of $10^6~$yr; at 1.5~Myr, the effective
temperature should be below 20~eV.

For J0720, there is an additional age estimate from kinematics
(\citealt{mzh03,kap04}; also Motch, these proceedings): tracing
back its proper motion, the most natural birthplace is the Trumpler 10
association; it would have left about $7\times10^5$~yr ago.  (Note
that a possible origin in the Scorpius OB associations about 1.5~Myr
ago has also been suggested, but the new parallax and proper motion we
derived [partly described in \S\ref{sec:d} below] make this less
likely.  Furthermore, for J1856, which is cooler than J0720, the
kinematic age of about $4\times10^5$~yr is not in doubt.)

Of course, the above discrepancy may simply mean that the
characteristic age is a poor estimate of the true age.  For breaking
with $\dot\nu\propto\nu^n$, where $\nu$ is the spin frequency and $n$
the so-called braking index (equal to 3 for magnetic dipole
radiation), the true age is $t=(1-P_0/P)^{n-1} (P/(n-1)\dot P)$.
Thus, one can obtain ages $t\ll\tau$ either if the initial spin period
$P_0$ is close to the current one (i.e., the neutron star was born
spinning slowly, but in no other systems is there evidence for
$P_0>1~$s), or if $n$ is substantially larger than~3.  Lyne (these
proceedings) presented evidence for values of $n$ substantially
different from~3, although most values were less, implying
characteristic ages that are longer than the true age, contrary to
what is required here.

As an alternative, we noted that one way of obtaining $P_0\simeq P$,
would be to have the neutron star undergo a phase in which it was
accreting, either from a companion (which later disappeared, e.g., in
a supernova explosion) or perhaps a fall-back disk such as that
discovered around the AXP 4U 0142+61 \citep{wck06}.  Intriguingly, the
equilibrium spin period, $P_{\rm{}eq}\approx
5{\rm\,s}\,(B/10^{13}{\rm\,G})^{6/7}(\dot M/\dot M_{\rm Edd})^{-3/7}$,
is roughly equal to the current observed periods for a magnetic field
of a few $10^{13}~$G and an accretion rate $\dot M$ close to the
Eddington rate $\dot M_{\rm Edd}$.

\begin{figure}
\begin{center}
\includegraphics[width=0.96\hsize]{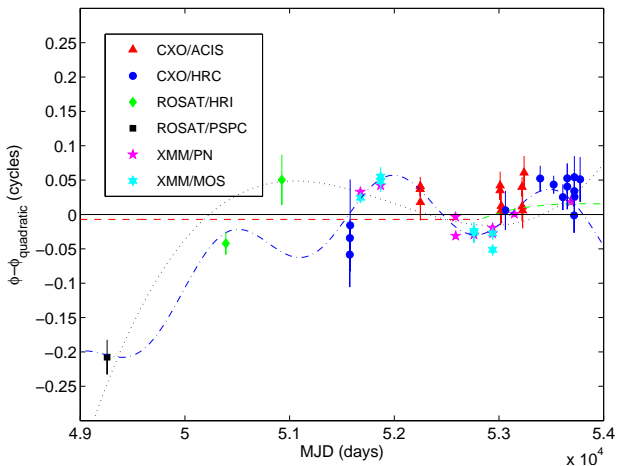}
\caption[]{Phase residuals for timing measurements for RX
J0720.4$-$3125 relative to the best-fit quadratic ($\ddot\nu=0$)
model.  The data include those from \citet{kvk05}, plus a number of
additional points from {\em Chandra} with LETG/HRC and one additional
point from {\em XMM-Newton} with EPIC-PN (all at ${\rm MJD}>53500$);
other EPIC-PN data exist (Haberl, these proceedings) but they are not
yet public.  The fit has $\chi^2_\nu=11.93$ and the the residuals have
${\rm rms}=0.39$~s.  We also show alternate fits: the best-fit cubic
model ($\ddot\nu\neq 0$; dotted, black curve), which has ${\rm
rms}=0.36$~s and $\chi^2_\nu=5.12$; the best-fit periodic model
(dot-dashed, blue), which has a period of 4.3~yr and ${\rm
rms}=0.23$~s and $\chi^2=2.35$; and a simple glitch model (dashed,
red) with the glitch occuring at MJD~52821, which does not
substantially improve the fit over the quadratic model.  Note that
there appears to be high-frequency residuals that are not fit by any
of these models, as can be seen by the disagreement between
simultaneous XMM observations with different instruments.  Whether
these are due to the energy dependence of the pulse profile or some
instrumental effect, we cannot say, but it suggests that caution
should be taken in interpreting the timing residuals.}
\label{fig:residuals}
\end{center}
\end{figure}

\begin{figure}
\begin{center}
\includegraphics[width=\hsize]{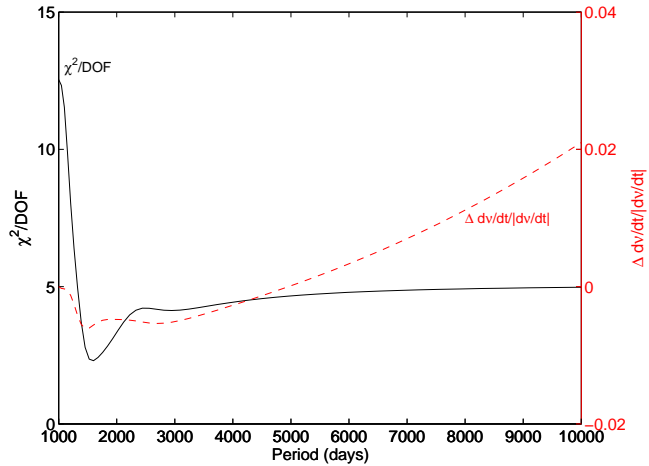}
\caption[]{Improvement in $\chi_\nu^2$ for a timing model for RX
J0720.4$-$3125 that includes a periodic component, as a function of
period, fit to the data shown in Fig.~\ref{fig:residuals}.  The
$\chi_\nu^2$ (solid, black curve) has its deepest minimum around
1580~days (4.3~yr); there may be a second minimum near 3000~days
(8.2~yr), close to the $\sim\!7~$yr period suggested by Haberl et
al. (2006), but with these data it is clearly less significant.  Also
shown is the fractional change in $\dot\nu$ for the different fits
(dashed, red curve); it changes by $<\!2$\% for the full range of
periods considered here, and thus does not influence the conclusions
regarding the magnetic field or characteristic age.}
\label{fig:periods}
\end{center}
\end{figure}

Finally, comparing the timing residuals, we find that for J0720, they
are $\sim\!$0.3~s, far larger than the measurement errors, while for
J1308, they are consistent with the measurement errors, at
$\sim\!$0.01~s.  The larger residuals for J0720 have been ascribed to
precession (\citealt{htdv+06}; also Haberl, these proceedings).  To
verify this, we tried including different terms in our timing model
(see Fig.~\ref{fig:residuals}), and we indeed find that adding a
periodic component improves the fit drastically, with the reduced
$\chi^2$ decreasing from $\chi^2_\nu=11.9$ to 2.4.  Trying different
periods, however, the best period appears to be 4.3~yr
(Fig.~\ref{fig:periods}), and not $\sim\!7$~yr as inferred from the
spectral changes (at 7~yr, the fit is better than quadratic, but not
that much different from a higher-order polynomial).  We caution,
though, that the timing residuals shown by J0720 are not exceptional:
they are in line with trends seen for radio pulsars and similar
apparent periodicities can be seen in the residuals of some of the
Anomalous X-ray Pulsars (Kaspi, these proceedings).

\section{Parallax Measurements}
\label{sec:d}

A parallax measurement for the brightest INS, J1856, was first
attempted by \citet{wal01}, using three observations with the
Planetary Camera onboard the {\em Hubble Space Telescope} ({\em HST});
the resulting parallax implied a distance of $\sim\!60$~pc.  The
measurement was tricky, however, and a much larger distance of 140~pc
was derived from the same observations by \cite{kvka02}; a larger
distance, of 117~pc, was also found by \citet{wl02}, who redid their
analysis and included a fourth PC observation.

In order to obtain more accurate distances, we have used the
High-Resolution Camera (HRC) of the Advanced Camera for Surveys (ACS)
onboard {\em HST}.  This camera is more sensitive than the PC and has
a smaller pixel scale, so that undersampling of the point-spread
function is much less of an issue.  Furthermore, across the pixels,
the sensitivity is more uniform, reducing the variability in the
point-spread function with pixel phase; as a result, much more
accurate astrometry can be done with ACS/HRC \citep{ak06}.  We
obtained images of J1856 and J0720 in the blue F475W band, visiting
each source eight times over two years.

For J1856, our analysis of the HRC data is virtually complete (Kaplan,
van Kerkwijk, \& Anderson, in preparation).  In order to obtain as
accurate a parallax as possible, we have taken into account the
parallactic motion of the background reference stars, by determining
their photometric parallaxes (assuming they are main-sequence stars;
for the less distant stars -- generally the brighter ones with strong
weight -- we confirm the photometric parallaxes astrometrically).
With that, from the HRC data alone, we determine a parallax
$\pi=6.2\pm0.6$~mas, corresponding to a distance
$d=161^{+18}_{-14}$~pc.  We are currently trying to improve the
measurement further by including the PC data.

For J0720, a first analysis of the HRC data has just been completed.
For this source, parallaxes of the background stars are much less
important, since it is at low Galactic latitude and most objects are
distant.  Our preliminary parallax is $\pi=3.0\pm1.0$~mas,
corresponding to a distance $d=330^{+170}_{-80}$~pc.

The factor two ratio in the distances to J1856 and J0720 is consistent
with what was expected by \citet{kvka02} under the zeroth-order
assumption that the optical flux for different sources scales as
$f_v\propto{}T(R/d)^2$, that the radii $R$ are similar, and that the
temperature $T$ in the region of the atmosphere emitting the optical
emission scales with the temperature determined from fits to the X-ray
spectrum.  The distances also compare well with the distances of
$135\pm25$ and $255\pm25~$pc inferred from the run of H{\,\sc i}
column density with distance (Posselt, these proceedings).

\begin{figure*}
\hbox to\hsize{%
\includegraphics[width=0.474\hsize]{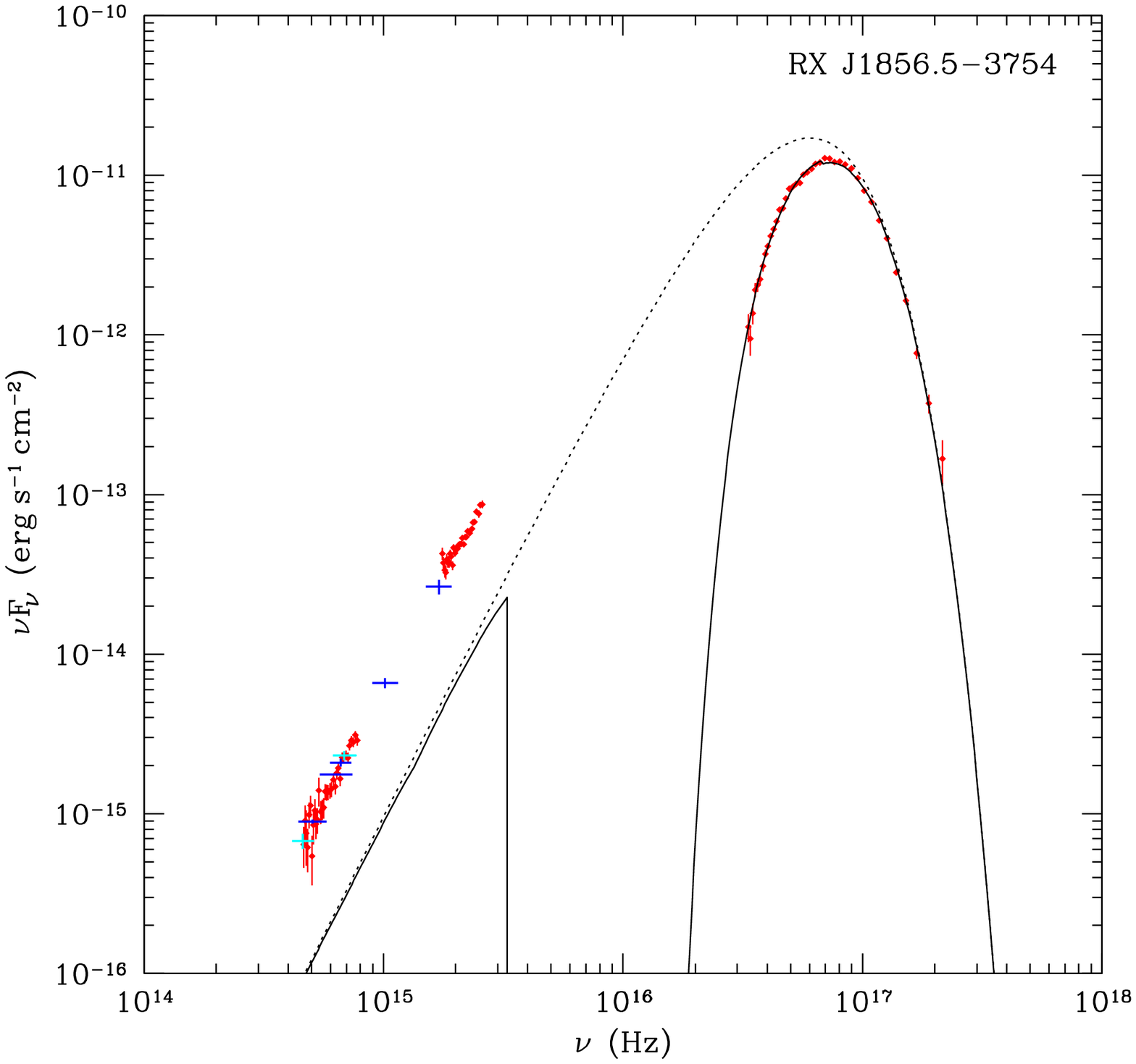}\hfill
\includegraphics[width=0.474\hsize]{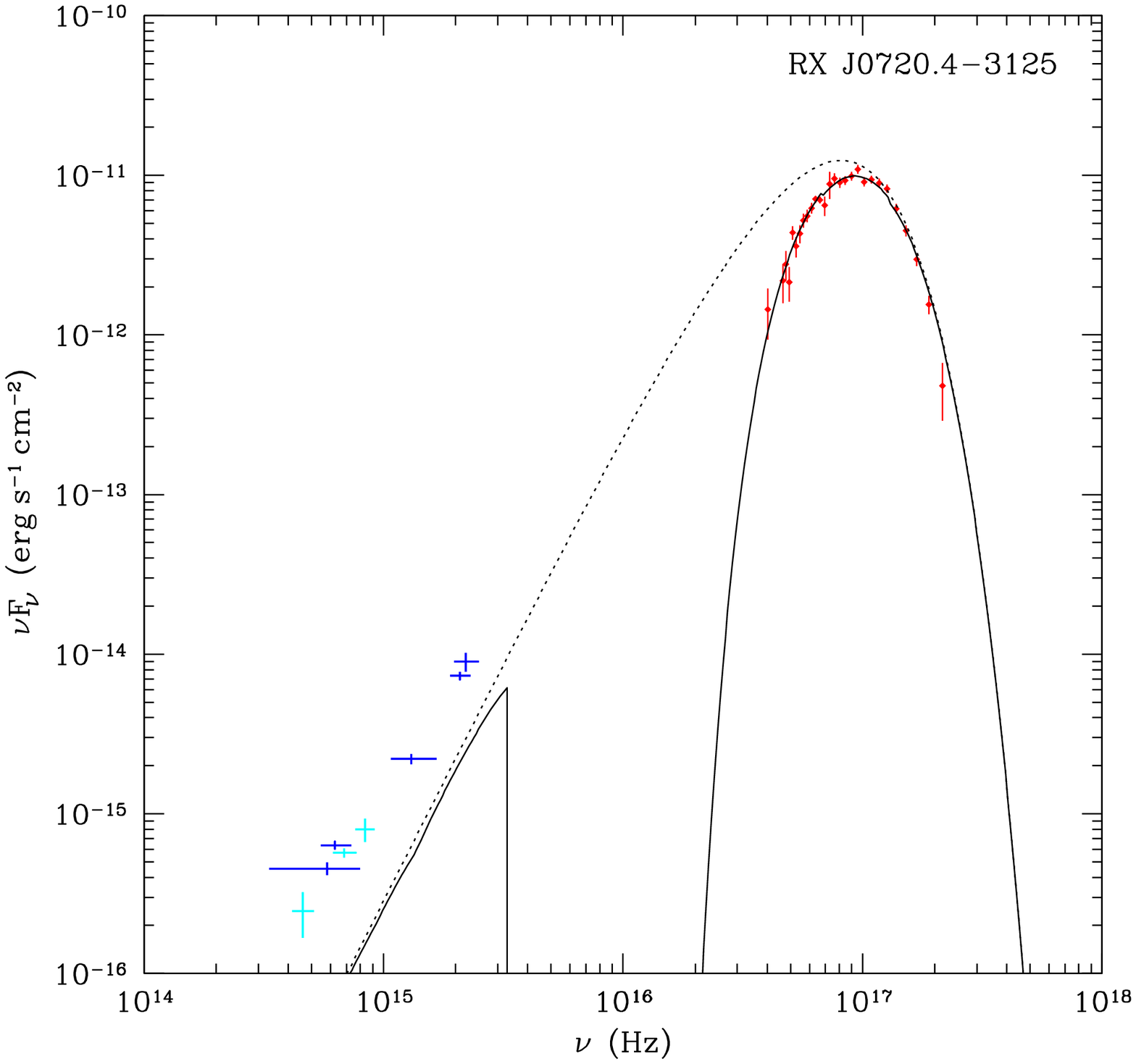}}
\caption{Spectral energy distributions for RX J1856.5$-$3754 (left)
  and RX J0720.4$-$3125 (right; with the X-ray spectrum from before
  the appearance of an absorption feature).  For both, the X-ray
  points are from LETG spectra, the dark blue points from {\em HST},
  and the cyan points from ground-based observations.  The optical and
  ultraviolet spectra for RX J1856.5$-$3754 are from VLT and {\em
  HST}, respectively.  The black, drawn curves represent the best-fit
  black-body models to the X-ray data; the dotted curves are the same
  model without interstellar extinction.} 
\label{fig:nfnj1856}\label{fig:nfnj0720}
\vspace*{0.5cm}
\hbox to\hsize{%
\includegraphics[width=0.474\hsize]{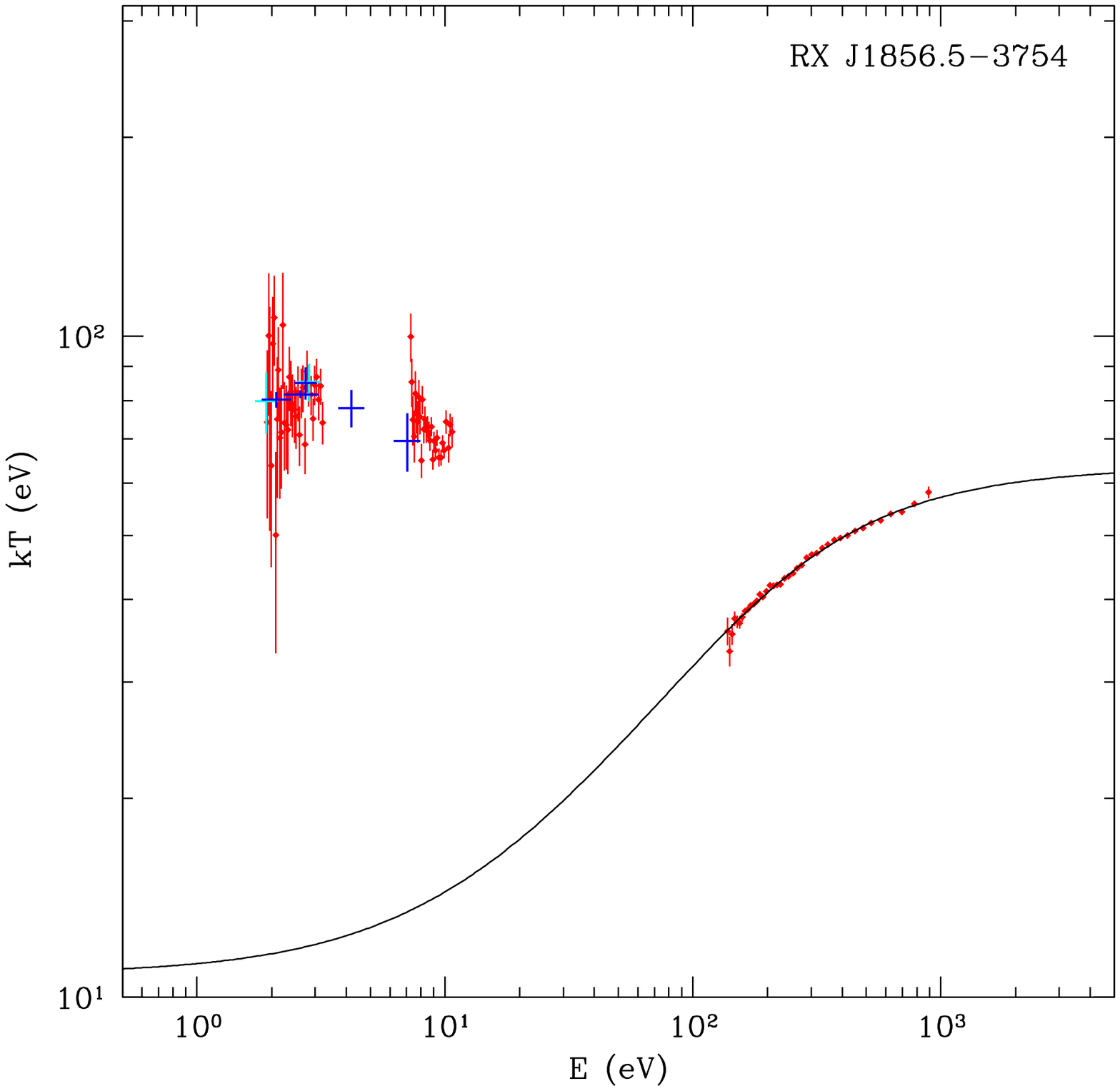}\hfill
\includegraphics[width=0.474\hsize]{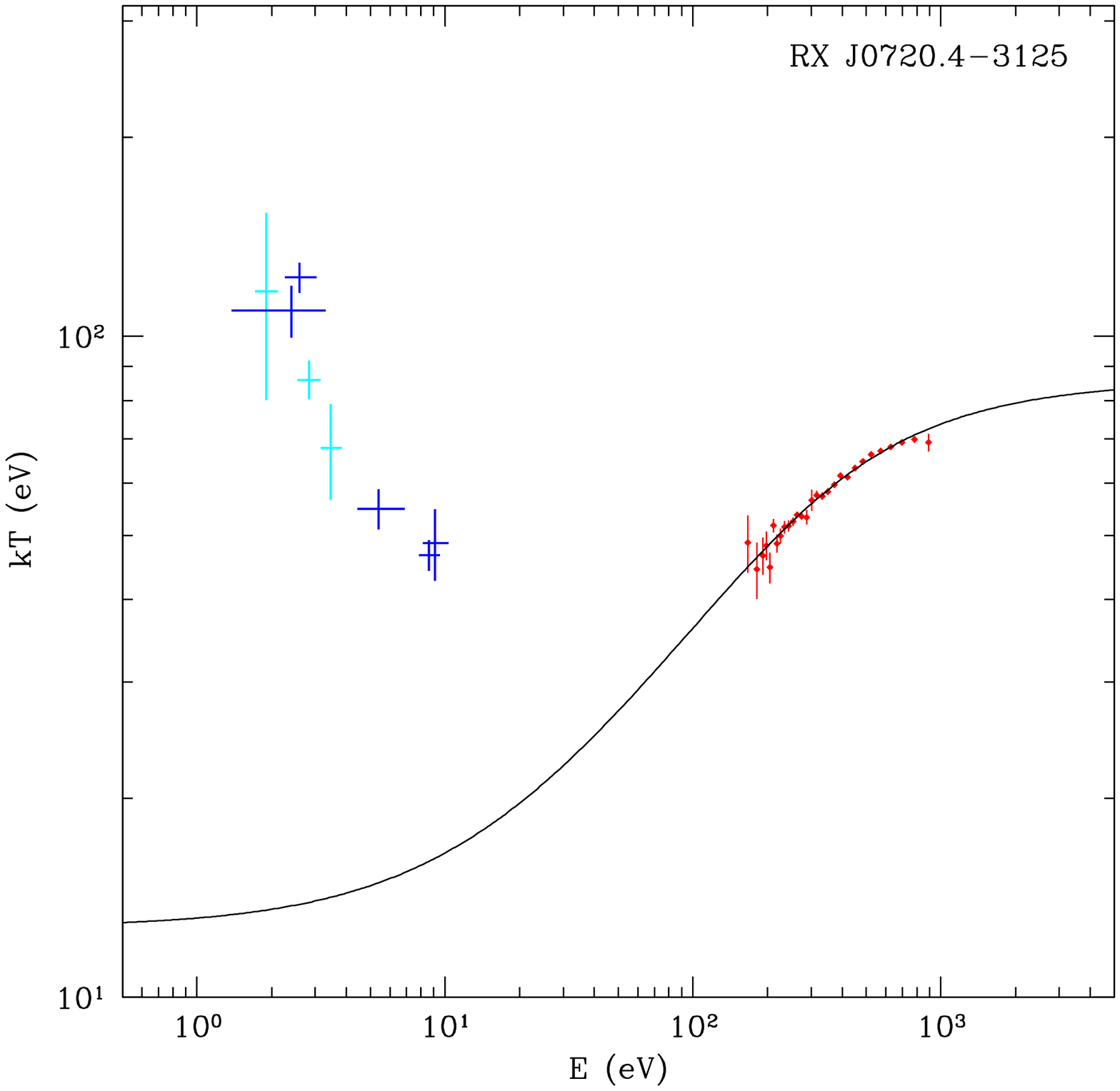}}
\caption{Brightness temperatures for RX J1856.5$-$3754 (left) and RX
  J0720.4$-$3125 (right), assuming our parallax measurements are
  correct and that the emission arises from a neutron star with
  radiation radius $R_\infty=14.7~$km (which is the value one obtains
  for $M=1.35~M_\odot$ and $R=12$~km).  The symbols and colours are as
  in Fig.~\ref{fig:nfnj1856} (the model curves are not at a constant
  temperature since the best-fit radiation radius is not equal to
  14.7~km).  One sees that the optical emission requires temperatures
  at least equal to those required for the X-ray emission.  Note,
  however, that for RX J0720.4$-$3125 it is not clear the emission is
  thermal.}
\label{fig:tbrj1856}\label{fig:tbrj0720}
\end{figure*}

With our distances, we can estimate the radii for the two sources.  We
start by simply using the black-body fit to the X-ray spectra.  For
J1856, one finds $kT=63$~eV and $R_\infty/d=0.0364{\rm~km\,pc^{-1}}$,
which, with our new distance, implies a radiation radius
$R_\infty\simeq6.5~$km.  This is smaller than a typical radius of a
neutron star, but this is not unexpected, for two reasons.  First, for
the most likely atmospheric compositions, the opacity decreases with
increasing frequency.  As a result, at X-ray energies one sees
relatively hot layers and a fit to the X-ray spectrum will thus
overestimate the effective temperature and underestimate the radius
\citep{pztn96}.  Second, the temperature distribution likely is not
uniform, in which case the area inferred from the X-ray emission would
simply correspond to that of the hotter parts.


In the above picture, one expects the optical emission to be in excess
of the extrapolation from the black-body fit, since it arises from a
cooler layer and from a larger area.  And indeed, the spectral energy
distribution, shown in Fig.~\ref{fig:nfnj1856}, shows an excess.  It
poses a possible problem, however, since the optical excess is a
factor~7, which implies a radiation radius of $R_{\infty,\rm opt} =
17(T_{\rm X}/T_{\rm opt})^2~$km (where $T_{\rm opt}$ and $T_{\rm X}$
are suitable averages of the temperatures of the optical and X-ray
emitting regions, respectively).  Given that one expects $T_{\rm opt}
< T_{\rm X}$, the optical emission thus seems to imply that the
radiation radius is quite a bit larger than 17~km.  Yet, for most
reasonable equations of state, a typical neutron star will have a
smaller radiation radius (e.g., \citealt{lp01}).

Of course, the above discrepancy may simply reflect our lack of
understanding of neutron star atmospheres in strong magnetic fields:
the temperature of the optical emission region may be larger than
expected.  In order to see what would be required, one can reverse the
process: assume that the neutron star has a `standard' mass and
radius, and calculate the brightness temperature at each energy
assuming that the emission originates from the whole surface.  In
Fig.~\ref{fig:tbrj1856}, we show the result for $R_\infty=14.7~$km
(which is the value one obtains for $M=1.35~M_\odot$ and $R=12~$km).
We see that this confirms the above reasoning: in order to produce the
optical emission, the temperature in the emission region has to exceed
70~eV, i.e., be higher than that in the X-ray emitting region.

For J0720, a fit to the X-ray spectrum from 2000 (i.e., before the
appearance of an absorption line) gives $kT=85.7~$eV and
$R_\infty/d=0.0170{\rm~km\,pc^{-1}}$.  Taking the distance at face
value, the implied radiation radius is 5.7~km, a little smaller than
that of J1856, but easily consistent within the 30\% uncertainty due
to the parallax measurement error.  Since the optical excess is
similar, the radiation radius for the optical emission is again large.
In this case, however, the optical emission does not follow a
Rayleigh-Jeans tail (\citealt{kvkm+03,mzh03};
Fig.~\ref{fig:nfnj0720}), and hence it is not clear that the emission
is from the surface.  This can also be seen from the brightness
temperatures (Fig.~\ref{fig:tbrj0720}), which is not constant in the
optical/ultraviolet range.

A further puzzle raised in comparing the sources, is that despite the
fact that the X-ray emission areas are rather similar, the pulsation
properties are very different: J0720 shows clear pulsations, with a
pulsed fraction of 11\% \citep{hmb+97}, while J1856 shows no
pulsations, to a limit of $\sim\!1$\% \citep{rgs02,bhn+03}.  This may
reflect differences in geometry; for isotropic emission from two
opposite magnetic poles, there is a fair range in parameters for which
no pulsations would be observed (e.g., \citealt{bel02}).  Of course,
the presence of the pulsations constitutes a warning about the
brightness temperatures shown in Fig.~\ref{fig:tbrj0720}: if the X-ray
emission does not arise from the whole surface, the true temperatures
will be higher than those shown.

\begin{figure}
\begin{center}
\includegraphics[width=\hsize]{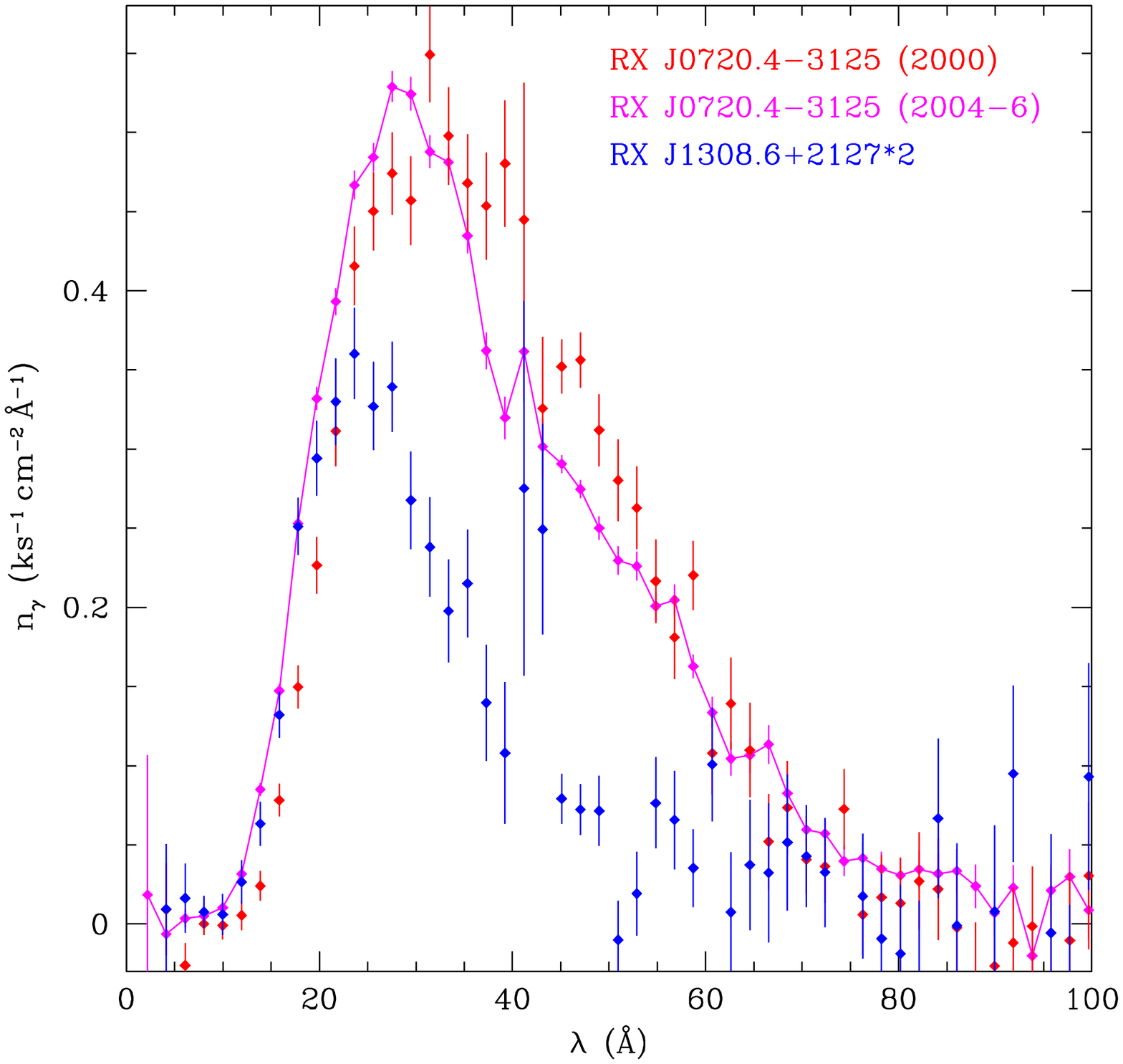}
\caption{Spectra of RX J0720.4$-$3125 and J1308.6+2127 taken with
LETG.  The magenta points connected with the line are the average of
spectra of RX J0720.4$-$3125 taken after 2004 (about 300\,ks).
Compared to the earlier spectrum (red points), the spectrum is harder
and has developed an absorption feature.  On the other hand, compared
to RX J1308.6+2127 (blue points, multiplied by two in order to match
at the short-wavelength side), the feature is rather weak, even though
the inferred temperature is similar (as is also clear from the good
match at short wavelengths), and so is the magnetic field inferred
from timing.}
\label{fig:letg}
\end{center}
\end{figure}

\section{LETG spectra}
\label{sec:spectra}

The study of the X-ray spectra of the INS has made great strides with
the advent of {\em Chandra} and {\em XMM-Newton}.  Both the CCD
instruments, in particular EPIC-PN, and the grating spectrometers LETG
and RGS have been used extensively.  Here, we focus on the grating
instruments (for the exciting results from EPIC-PN, see Haberl, these
proceedings).  We will only discuss results from LETG, since that
instrument covers the full range of energies at which INS emit and
since the calibration of the RGS at longer wavelengths has been rather
problematic.\footnote{The problems appear to be largely solved with
the 2006 June 30 release of the Scientific Analysis Software.}

So far, three sources have been observed with LETG.  By far the best
spectrum is of J1856, taken using 500\,ks of director's discretionary
time.  Unfortunately, and puzzlingly, the spectrum appears completely
featureless, and is well described by a black-body model
\citep{bzn+01,bhn+03,dmd+02,br02}.  The second brightest source, J0720
has been studied extensively as well.  A first spectrum was taken in
2000 \citep{kvkm+03}, when the spectrum was featureless, and a second
in 2004 \citep{vdvmv04}, to confirm the change in spectrum discovered
with RGS by \citet{dvvmv04}.  During 2005, the source was regularly
monitored by us, for 300\,ks total, in order to look for further
spectral changes and to see if more than one spectral feature was
present.  Finally, a 100\,ks observation was made of J1308 in
guaranteed time by the MPA group.

In Fig.~\ref{fig:letg}, we show the LETG spectra of J0720 and J1308.
For the former, we show separately the `before' spectrum (from 2000,
before the appearance of an absorption feature) and the `after'
spectrum (the average of all spectra taken after the change; within
our statistics, the individual spectra do not differ).

The comparison of the three spectra raises a number of questions.
First, for J0720, how can a relatively small change in temperature
(from $\sim\!80$ to $90~$eV) cause the appearance of a pronounced
absorption feature? Second, independent of the state of J0720, why is
the absorption feature much less strong than that in J1308, despite
the fact that their temperatures and magnetic field strengths (from
timing, \S\ref{sec:timing}) are rather similar?

Considering first the change in J0720, the simplest possibility would
be that the increase in temperature corresponds to a relatively large
change in the ionisation and/or dissociation balance.  If so, this
might give a clue to the corresponding energies of the matter in the
atmosphere.  The alternative would be that the region that heated up
(or appeared in view) has either a different composition or a greatly
different magnetic field strength compared to the regions that
dominated the spectrum before the change.  Neither possibility seems
particularly appealing.


Whatever the physical reason for the appearance of the absorption
feature, another more basic question is whether the change corresponds
to a global change in the properties, or whether, instead, only a
fraction of the surface changed or if ones viewpoint changed.  A
global change might occur if, e.g., heat was deposited deep inside the
neutron star.  In contrast, a change in a limited area would be
expected if heat were deposited near the surface (with the affected
area perhaps increasing in size with time), or if the source were
precessing and different regions came into view
(\citealt{dvvmv04,htdv+06}; Haberl, Zane, these proceedings).

Of course, if only part of the area that we see changed, then the
average `after' spectrum we currently observe contains a contribution
from the unchanged parts of the surface, i.e., from the cooler,
featureless `before' spectrum .  Hence, the spectrum from the changed
part should be hotter and should have a stronger line than one would
infer from the average.  We can set an upper limit to the contribution
from the `before' spectrum by requiring that it does not exceed the
`after' spectrum at any wavelength.  From Fig.~\ref{fig:letg}, one
sees that the limit is about 70\%, set by the 35--40\,\AA\ region.

The above could solve the second question: it might well be that after
the change, the parts of the surface of J0720 that show an absorption
feature in their spectrum, have a line as strong as that observed in
J1308.  It only appears weaker in the `after' spectrum because it is
diluted by the featureless emission from the unchanged parts of the
surface.  So, just a single question may be left: how can a
neutron-star atmosphere, with presumably the same magnetic field
strength and the same composition, and with only a modest, $\Delta
T/T<0.2$ temperature increase, emit such different spectra?

\section{Strongly magnetised atmospheres}
\label{sec:atmosphere}

In interpreting the spectra, a major uncertainty is the composition.
For a single source, this may be difficult to determine uniquely, but
one can hope to make progress by treating the INS as an ensemble:
ideally, it should be possible to understand the features (or lack
thereof) in all INS with a single composition, appealing only to
differences in temperature and magnetic field strength (constrained by
observations where possible), which might lead to different ionisation
states being dominant, and possibly the formation of molecules or even
a condensate.  Here, we discuss only the possibilities of hydrogen or
helium atmospheres.  For completeness, we note that gaseous
atmospheres composed of heavier elements appear to be excluded by the
lack of large numbers of features.  Condensated from heavier elements
are also being considered seriously (Pons, Ho, these proceedings), and
detailed theoretical calculations are being carried out to determine
at what magnetic field strength condensates can form (\citealt{ml06};
Lai, these proceedings).

\subsection{Hydrogen}

\begin{figure}
\begin{center}
\includegraphics[width=\hsize]{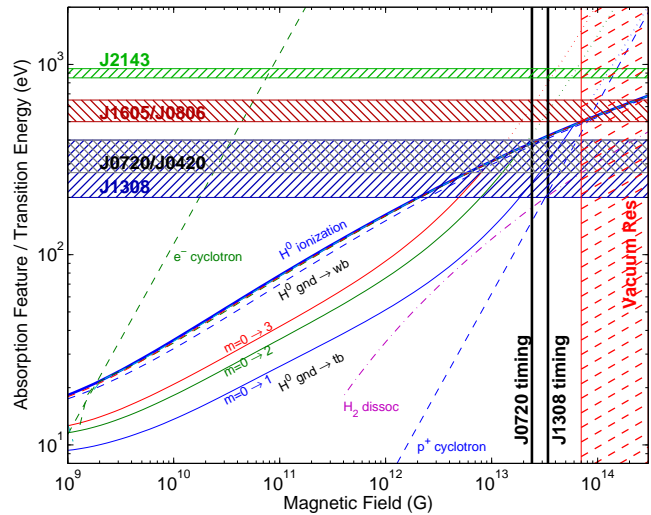}
\caption{Energy versus magnetic field for electron and proton
cyclotron, ground state to tightly bound (tb) or weakly bound (wb)
states in neutral hydrogen, hydrogen ionisation, and molecular H$_2$
dissociation \citep{hlpc03,lai01,pot98}.  For fields above
$\sim\!10^{14}\,$G (hatched region to the right), features may be
washed out due to the effects of vacuum resonance mode conversion
\citep{lh03}.  The hatched bands show the energies of the main
absorption features (corrected for gravitational redshift) for the
INS, and the two vertical lines indicate the dipole field strengths
determined by timing.  The sources are labeled with short-hand
notation as in the text (see end of \S1).  References for the energies
are: J0420, \citet{hmz+04}; J0720, \citet{hztb04}; J0806,
\citet{hmz+04}; J1308, \citet{hsh+03}; J1605, \citet{vkkd+04}; J2143,
\citet{zct+05}.}
\label{fig:h}
\end{center}
\end{figure}

The presence of a hydrogen atmosphere has often been considered by
default, since if any hydrogen is present, gravitational settling will
ensure it floats to the surface.  Typically, it has been assumed the
hydrogen is fully ionised, and spectral features have been interpreted
as proton cyclotron lines.  In strong magnetic fields, however, the
binding energies of atoms increase (for a review, \citealt{lai01};
Potekhin, these proceedings; see also Fig.~\ref{fig:h}), and for
temperatures and fields appropriate for INS, a fraction of up to 10\%
of neutral hydrogen will be present \citep{pcs99}.  From initial
model-atmosphere calculations that take the presence of neutral
hydrogen into account (\citealt{hlpc03}, see their Fig.~3), it is
clear that, e.g., at $10^6~$K and $10^{13}~$G, the lines from neutral
hydrogen have larger equivalent width than the proton cyclotron line
(they are less deep but much wider, due to the so-called motional
Stark effect; \citealt{pm93,pp97}); at lower temperatures or stronger
magnetic fields, the fraction of neutral hydrogen increases and hence
the difference should be even larger.  In general, it is worth
stressing that the features are very strong: they may not appear so on
the logarithmic scale typically used, but they have depths depth often
exceeding 50\%, similar to what is observed for J1308.

Below, we first discuss whether the energies of the main features
observed in the INS can be reproduced by a strongly magnetised
atmosphere, and then turn to two possible problems: harmonically
spaced lines found recently, and the featureless, black-body like
spectra shown by some INS.  We do not include the optical excess among
these problems, since currently it is not clear any model makes
reliable predictions for the optical emission: at a few $10^{13}~$G,
the plasma frequency exceeds the frequencies of optical photons, and
the models do not take into account the resulting significant
deviations of the refractive index from unity (\citealt{val06}; see
\citealt{ks04} for a discussion of possible effects in the context of
cool white-dwarf atmospheres composed of helium).

\subsubsection{Line energies}

In Fig.~\ref{fig:h}, we show the energies for features that might be
produced in a hydrogen atmosphere: the electron and proton cyclotron
lines, and the bound-bound and bound-free transitions of neutral
hydrogen (relative to the ground state).  Also shown are the
approximate energies of the main features that have been detected in
the various INS (corrected for an assumed gravitational redshift of
0.3)

In the figure, thick vertical lines indicate the two magnetic field
strengths inferred from timing~(\S\ref{sec:timing}).  From those, it
follows that if J0720 and J1308 have hydrogen atmospheres, the
features are most likely due to the transition from the ground state
to the first excited tightly bound state of neutral hydrogen, perhaps
in combination with the proton cyclotron line.  As argued in
\S\ref{sec:spectra} above, the line in J0720 might be weaker than that
in J1308 because the emission from part of its surface is featureless
(which in itself is problematic; we return to this below).
Alternatively, the line in J0720 might be weaker because it is to the
second excited tightly bound state \citep{vkkd+04}.

If the above is correct, the feature in J0420 likely has the same
origin and thus its field should also be a few $10^{13}~$G.  The
features in J1605 and J0806 could result from the same transitions or
from the ionisation edge, but in either case the implied magnetic
field strength is higher, close to~$10^{14}~$G.  For J1605,
\citet{vkkd+04} noted that the line was substantially weaker than that
of J1308, and they suggested this might be due to the effect of vacuum
resonance mode conversion, which for fields in excess of
$\sim\!7\times10^{13}~$G tends to weaken features (see \citealt{hl03};
also Lai, these proceedings).  Finally, for J2143, the line energy of
0.7~keV is substantially higher than what is observed for all other
sources, and for any transition in neutral hydrogen, the upper state
is auto-ionising: it is at an energy level that is higher than the
continuum from the ground state.  It is thus not clear whether the
line could be due to neutral hydrogen.  Instead, it might be due to
the proton cyclotron line in a field of just over $10^{14}~$G.  For
these field strengths, the feature should be strongly weakened by
vacuum resonance mode conversion (but not necessarily disappear;
e.g., \citealt{hl04,val06}); qualitatively, this is consistent
with the rather modest observed strength \cite{zct+05}.

\subsubsection{Possible problem 1: Harmonically spaced lines}

At the conference, evidence for harmonically spaced absorption lines
was presented for three INS.  For J1605, Haberl (these proceedings)
found that apart from the line at 0.40~keV discovered by
\citet{vkkd+04}, the EPIC-PN data show a significant feature at
0.78~keV, i.e., at an energy that is in a 1:2 ratio with that of the
stronger line.  Furthermore, a third feature at 0.59~keV could be
present, consistent with energies in a 2:3:4 ratio.  For J1308,
Schwope et al.\ (these proceedings) presented evidence that the single
strong feature originally found at 0.3~keV or less by \citet{hsh+03},
could be composed of two features, at 0.23 and 0.46~keV, i.e., again
harmonically spaced.  Finally, for J0806, the single feature at
0.43~keV found by \citet{hmz+04} may again be better described by two
features at 0.30 and 0.60~keV (Haberl, these proceedings).

It would appear tempting to interpret these features as cyclotron
lines, since those naturally have harmonic energy ratios.  It is
difficult, however, to see how this could be possible for proton
cyclotron lines, since the harmonics are expected to be exceedingly
weak: the oscillator strength for the harmonic would be a factor
$E/m_{\rm p}c^2$ weaker than that for the fundamental.

Instead, as mentioned in a discussion with George Pavlov, Joachim
Tr\"umper, and Frank Haberl at the meeting, a different solution may
be suggested by the behaviour of the transitions of neutral hydrogen.
As can be seen in Fig.~\ref{fig:h}, for any transition, above a
certain magnetic field strength, the transition energy starts to
become proportional to the proton cyclotron energy.  As a result, at
sufficiently strong magnetic field, the transitions become
harmonically related.  A possible problem, however, is that in this
situation, the upper level of the transition is an auto-ionising
state, i.e., it has an energy in excess of the continuum energy
relative to the ground state.  It will still lead to some additional
opacity, but at present it is not clear whether this is sufficient.
Fortunately, there is one prediction: for J1605, it would not be
possible to explain the spectrum if there are really three features in
a 2:3:4 ratio, without a strong corresponding `1'; thus, the
prediction is that upon further analysis, the 0.59~keV feature will
disappear.

Finally, we note in this context that it will be worth checking
carefully that for J2143, the 0.7~keV feature observed is in fact not
a `harmonic.'  From the present fits by \citet{zct+05}, a rather high
$N_{\rm H}$ is inferred, and this could perhaps be an artefact of a
strong absorption feature at $\sim\!0.3~$keV? (From initial attempts,
this appears unlikely; Cropper, 2006, pers.\ comm.)

\subsubsection{Possible problem 2: Featureless black-body spectra}

Perhaps the most severe problem with the idea that the INS have pure
hydrogen atmospheres is that the spectra of J1856 and J0720 (before
the change) are featureless and well represented by black-body
emission.  For J1856, perhaps no features are expected, since its
magnetic field strength, as inferred from the bow-shock shaped
H$\alpha$ nebula around the source \citep{vkk01b,kvka02}, is below
$10^{13}~$G, in which case all features may be below the observed band
(Fig.~\ref{fig:h}), but for J0720 this explanation is not possible.
Furthermore, for a mostly ionised atmosphere, the spectrum is expected
to have a hard tail, unlike the observed exponential, Wien-like shape,
since the free-free opacity decreases with increasing energy.

There are several possible solutions.  First, there could be a reason
for the opacity to be much greyer than currently estimated, so that
the emission at all wavelengths originates from layers at similar
depths and thus with similar temperature.  The extreme version of
this, discussed in detail by Pons and Ho (these proceedings), is that
the two sources have a condensed surface (with possibly only a thin
hydrogen layer on top).  A less extreme version might be that the
atmosphere does not contain just ionised and neutral hydrogen, but
also molecules, which might have so many transitions that the opacity
becomes effectively grey.  Hydrogen molecules do indeed have a higher
binding energy than hydrogen atoms, but the dissociation energy is
only around 0.2~keV for a few $10^{13}~$G (\citealt{lai01}; see
Fig.~\ref{fig:h}).  With temperatures only a factor two smaller, the
abundance should be very small (as indeed found by, e.g.,
\citealt{pcs99}; see their Fig.~7).  Nevertheless, it may be worthwhile
verifying this, making sure that the abundance and the resulting
opacity are indeed negligible.

A possible alternative way to produce spectra resembling black bodies
is by making the temperature profile in the atmosphere shallower,
closer to isothermal.  While this certainly appears ad hoc and likely
would require significant fine-tuning, there is evidence for active
magnetospheres: for J1856, the H$\alpha$ nebula provides evidence of a
pulsar wind, and for J0720, the optical emission appears to be partly
non-thermal.  If there is an active magnetosphere, some particles
might hit the atmosphere, leading to additional heating; at the right
locations, this could lead to rather different emergent spectra (e.g.,
\citealt{gbr02}).

At present, none of the above explanations seem satisfactory.  Also,
none provide an easy explanation for why some sources have featureless
spectra while others have not (or why it would change).  Perhaps the
first parameter to consider would be the overall temperature, since
J1856 is cool and the appearance of the absorption feature in J0720
was accompanied by a temperature increase.  The increase was only
small ($\Delta T/T<0.2$), however, and furthermore, an absorption
feature does appear to be present in the coolest INS, J0420
($kT\simeq45~$eV, \citealt{hmz+04}).

\subsection{Helium}

Above, we stated that if any hydrogen were present, it would float to
the top.  Recently, it has been questioned, however, whether an outer
hydrogen envelope can survive \citep{cb04,cab04}.  The reason this is
not certain is that some hydrogen will diffuse down and reach
underlying Carbon or Oxygen layers, where, if the temperature is
right, it will be burned.  Indeed, \citet{cb04} find that all of the
hydrogen can be burned in the first $10^5~$yr of a neutron star's
life, in which case an atmosphere composed of helium might be left
(unless hydrogen is replenished, as could happen due to spallation by
relativistic particles from the magnetosphere, or very low levels of
accretion).

Partly inspired by this possibility, \cite{pb05} calculated properties
of singly-ionised helium in strong magnetic fields.  For a few
$10^{13}~$G, the transition energies are again in the range that
features are observed in the INS, and hence it seems worthwhile to try
to do a similar analysis as done above for hydrogen.  From a very
rough first attempt at producing model atmospheres (done by Kaya Mori
and Wynn Ho), including neutral helium, it seems that, like for
hydrogen, the features will be very strong.  Typically, however, more
than one very strong feature should be present, which appears to be in
conflict with what is observed.  The picture is currently incomplete,
however, since molecules have not yet been considered, while for
helium the binding energy of, e.g., He$_2^+$ is sufficiently high that
it may well be present (a detailed calculation is tricky, since one
has to have a decent estimate of the number of possible rotational and
vibrational states).

\section{Discussion and future prospects}
\label{sec:discussion}

Of the four main parameters mentioned in the introduction that
determine the properties of the thermal emission from INS, we now
appear to have reasonable handles on three: the shapes of the X-ray
spectra indicate temperatures around $10^6~$K, period derivatives
imply magnetic field strengths of a few $10^{13}~$G, and parallax
measurements show that a fair fraction of the surface is emitting
X-ray radiation.

The main unknown appears to be the composition.  We found that the
energies of the observed absorption features can be matched fairly
easily for hydrogen atmospheres.  However, reproducing the smooth,
featureless spectra of some INS, and the Wien-like high-energy side of
the X-ray spectra in general, appears problematic, nor is it clear how
the spectrum of J0720 could change from featureless to one that has an
absorption line.

Fortunately, it should soon become clear whether these issues are real
problems or not, since great progress is being made in constructing
more reliable strongly magnetised hydrogen model atmospheres (Lai,
Potekhin, these proceedings).  From \S\ref{sec:atmosphere}, it seems
particularly important to include in full detail transitions to the
auto-ionising levels, verify that all sources of opacity, including
from (traces of) molecules are included, and check the influence, in
particular on the temperature profile, of high-density effects and
vacuum resonance mode conversion.  At the same time, it would seem
worthwhile to consider atmospheres of other elements; for the INS, He
might be most relevant, but it would be good to check heavier elements
as well, since these may cause the absorption features seen in 1E
1207.4$-$5209 (\citealt{hm02}).

From the observational side, the easiest route to further progress
would appear to be timing.  With further estimates of the magnetic
fields, one can test the predictions based on hydrogen atmospheres,
that J0420 has a field about as strong as that of J0720 and J1308,
J0806 a stronger one, approaching $10^{14}~$G, and J2143 the
strongest, in excess of $10^{14}~$G.

For the X-ray spectra, further monitoring is useful, but perhaps the
largest advance will come from the unified analysis of all sources,
which allows one to exclude instrumental effects.  This is already
well underway for the EPIC-PN data (Haberl, these proceedings), and
similar studies of the LETG and RGS data should prove fruitful.  As
present, first steps are being taken in detailed modelling of the
phase-resolved spectra (Haberl, Zane, these proceedings), and this
should help obtain stronger constraints on the thermal distribution
over the surface.

Finally, in the optical-ultraviolet regime, it would be good to
complete the census of the sources, and obtain at least rough spectral
energy distributions, to determine whether the emission is thermal, or
whether there are non-thermal components.  For sources that are
sufficiently bright, proper motion measurements can help determine
true ages and parallax measurements can help determine distances.

\begin{acknowledgements}
It is a pleasure to thank Kaya Mori, Wynn Ho, George Pavlov, and Dong
Lai for enlightening discussions about the physics of strongly
magnetised, dense atmospheres, and Jay Anderson for his continuing
patience and collaboration in trying to extract the best possible
astrometry from {\em HST} images.  We acknowledge financial support
through a guest observer grant from NASA, as well as through
individual grants from NSERC (MHvK) and a Pappalardo fellowship
(DLK). 
\end{acknowledgements}

\bibliographystyle{apj}
\bibliography{ins}

\end{document}